\documentclass[11pt]{article}

\usepackage{amsmath}
\usepackage{slashed}

\title{The Multiverse and Particle Physics\footnote{Invited review for Annual Reviews of Nuclear and Particle Science}}

\author{John F. Donoghue \\ \\
Department of Physics,\\ University of Massachusetts, \\Amherst, MA, USA, 01003}
\date{January 3, 2016}

\begin{titlepage}
\begin{document}
\maketitle

\begin{abstract}
The possibility of fundamental theories with very many ground states, each with different physical parameters, changes
the way that we approach the major questions of particle physics. Most importantly, it raises the possibility that these different parameters
could be realised in different domains in the larger universe. In this review, we survey
the motivations for the multiverse and impact of the idea of the multiverse on the search for new physics beyond the Standard Model.
\end{abstract}


\end{titlepage}

\tableofcontents

\eject
\section{THE MULTIVERSE}

\subsection{Sociology}

When dealing with the fundamental interactions, one of the key science questions at the moment is whether the underlying theory of Nature has a unique ground state, a few possible ground states, or very many ground states. As we will see, this simple question leads into the idea of the multiverse - multiple domains in the universe with different properties - as well as anthropic reasoning.

We may as well face up front the issue that the multiverse and anthropics engenders strong opinions among individual scientists as well as difficult questions about the process of doing science in a multiverse. A common attitude is that work on mulitverse and anthropic considerations is distasteful and not even scientific. However, at the other extreme one has to recognize that not everything that relates to the multiverse is really good science. The community is gradually trying to understand the the implications of the possibility of a multiverse.

The point of view taken in this review is that it would be unscientific if we did not take the idea of a multiverse seriously as it is a real physical possibility. Exploration of multiverse ideas provides new insights into the primary puzzles of fundamental physics and often changes our approach to searches for new physics. So in this regard, it is a valuable motivator of fundamental theories. However, while we may eventually be able to test some predictions of a concrete multiverse theory, it is also true that the job of testing many aspects of such a theory is likely out of reach. We may run into a fundamental barrier to what we can know. However, the possible existence of multiverse theories is exciting at the moment and deserves more investigation.

Let us start with a situation where reasoning similar to that of multiverse theories is accepted without controversy.

\subsection{Analogy: The Earth-Sun distance} In trying to understand the nature of the solar system, it is natural to ask about the laws that govern the distance of the Earth and other planets from the Sun. In his {\em Mysterium Cosmographicum} \cite{Kepler}, Johannes Kepler proposed an elegant geometric explanation - that the radii correspond to the size of the five Platonic solids, inscribed in spheres and nested inside each other. The Earth would correspond to the dodecahedron, outside of the icosahedron, itself outside of the octahedron. Initial evidence was consistent with this structure, although eventually it was clear that the model does not work in detail. Despite its elegance, the model was incorrect. Should we continue to look for a fundamental law that predicts the Earth-Sun distance?

This science question can also take personal turn. Given our understanding of chemistry and biology, it also becomes clear that life as we know it would not be possible on Earth if the distance to the Sun were significantly different. Many of the other planets appear have physical properties that are not conducive to life.  What is the design that placed the Earth at just the right distance from the Sun in order to sustain life?

Of course, we now understand that the Earth's distance, and many other properties of the Solar System, are not directly predictable from fundamental laws. They  occur as an accident of the past history of the matter that collapsed gravitationally to form the Sun and planets. There are other stars with different luminosities and other solar systems with planets that have other distances from their star. And out of the multiplicity of all these solar systems and planets, we should not be surprised to find that we find ourselves on a viable planet. Given the conditions for the existence of life as we know it, only a subset of the many planets would be suitable and we could only find ourselves on one of these.

In this setting, it is not a useful scientific endeavor to search for a fundamental law uniquely predicting the Earth-Sun distance. There is no such law of physics that we know of - we are not that special. However, there could be a valid scientific attempt to predict the distribution of distances of planets from their stars, based on primordial matter distributions and gravitational force laws. There can be a  more fundamental theory that describes the mechanism of planet formation and this can make predictions for this distribution or ``measure'' in a statistical sense, given the primordial matter distribution.  From our own solar system we could get some constraints on this measure, although the statistics would be weak because we have only a small number of examples. If we explore other solar systems, we could potentially test the fundamental theory with more precision.

Kepler's quest to understand the planets mirrors our own attempt to understand the parameters of the Standard Model, such as quark and lepton masses and couplings. We would like a fundamental theory that rigidly predicts these, preferably from some elegant construction. We also have a personal connection in that the parameters that we measure appear remarkably fine-tuned for atoms, nuclei, and life. There have been many attempts to create predictive theories for all the masses and couplings, but they have yet not been successful in detail. Should we continue to search for such a theory, one which applies universally everywhere in spacetime? This is the standard assumption of particle physics and of course continues to be worthwhile. But, perhaps the more fundamental theory allows these parameters to take on different values, and perhaps these different values can be realized differently in various domains of the universe. This would be the multiverse solution. The analogy with Kepler's quest indicates that we at least need to explore this possibility.

\subsection{Conditions for a multiverse}

The Standard Model \cite{DSM, Langacker, GT2, MattS}, with the measured values for the parameters, has a unique ground state. It is found by looking for the minimum energy state of the Higgs potential. Somewhat more generally, the structure of the Standard Model could allow two ground states if the $\mu^2 H^\dagger H$ term in the Higgs potential were allowed to take on different signs. These ground states, one with an unbroken symmetry and the other with the spontaneous symmetry breaking, have vastly different properties \cite{agrawal, agrawal2}. If one considers Grand Unified Theories with more complicated scalar potentials \cite{GUTS}, such theories most often contain multiple ground states, again with vastly different properties. While this was enough to start multiverse thinking \cite{Linde:2015edk} it is not sufficient for most present discussions of the multiverse.

For the purposes of this review we will take the phrase multiverse to imply that there are multiple domains within the larger universe, each with different properties such as different values of the physical parameters, and perhaps even different gauge structures. These domains are contiguous, although generally domain walls would exist between them. We are not considering here alternate uses of the phrase, for example as variations of the many-worlds interpretation of quantum mechanics, which is a quite different beast. (Some other reviews with a similar focus can be found in \cite{Linde:2015edk, Carr:2007zzb, Donoghue:2007zz, Schellekens:2013bpa, Hogan, Cahn}. )

How many ground states are needed for a functioning multiverse? The requirement is that there be enough states that the parameter space is populated so densely that it is likely that one or more of the states looks similar to our own. In the next section we will see that this is a reasonably tight constraint, so that the original theory must possess very, very many ground states. If we include the cosmological constant in the constraints and the parameter distribution is relatively flat, this amounts to at least $10^{60}$ ground states, perhaps more than $10^{120}$. This is not for the faint of heart. It requires a qualitative change from the theories that we normally study, with at most a few ground states. However, it is a physical possibility. We most often assume that the parameter space is discrete, although it is possible for it to be continuous.

In the above, I use the concept of a parameter space. The idea is that if the parameters of the theory are not uniquely determined, they will have a possible range of values. For example, we see that the Yukawa couplings of the Standard Model range at least\footnote{The observed range could be larger if the small neutrino masses come from small Yukawa couplings.} from that of the electron with a coupling of $3 \times 10^{-6}$ up to the top's coupling of essentially $1$. So we would expect that the range of each of the Yukawa couplings is about a spread of unity on a linear scale. The gauge coupling constants in the Standard Model, at the weak scale, ranges at least from the QED coupling of $\alpha(M_Z) =1/128$ up to the strong interactions $\alpha_s(M_Z)= 0.11$. The union of all these ranges is the overall multi-dimensional parameter space. The total size is a bit fuzzy in the absence of a specific fundamental theory. But more importantly the shape/distribution/measure of the parameter space is unknown. Are the parameters arranged evenly on a linear scale, or perhaps on a logarithmic scale, or something else? Presumably different multiverse theories would have different measures. This is relevant for the discussion of the number of states needed. However, to address this realistically needs a specific underlying theory. We are not there yet.

The other physical requirement for the multiverse is a mechanism to populate the different ground states to form a multiverse of different domains with at least a reason why our domain looks roughly uniform and isotropic. At the moment this actually seems like the easier requirement, as long as inflation or something like it is at work in our domain. Inflation takes a small patch of the very early universe and spreads it out to a larger spatial extent than the limits of our observable universe. Inflation also allows quantum fluctuations, and potentially tunnelling events, that can populate different values of the fields.

Indeed, it was in the theory of inflation that the idea of a multiverse first arose \cite{Linde:2015edk}. Inflation driven by a scalar field will have a multiverse character of sorts, in that some other regions of the universe will still be inflating and will inflate at different amounts. Our domain will have stopped inflating and gone through reheating, while others are not there yet. However, the earliest inflationary descriptions did not contain a key ingredient for what we now consider as a key part of the multiverse - the very large number of ground states that are needed for the low energy theory. Even if inflation is still taking place elsewhere in the universe, there is a separate question of whether those domains which have stopped inflating will settle down into a unique ground state or into very many possibilities. However, if the multiple grounds states do exist in the fundamental theory, then inflation become a useful ingredient as it can easily be invoked to populate these states in different domains of the greater universe. Theorists can use inflation to cook up initial conditions that populate different ground states in different locations, yet allow our domain to appear smooth. For the particle physics discussion below, we will simply assume that this has happened in the early universe.

\section{FINE-TUNING: MOTIVATION FROM THE BOTTOM UP}

After decades of hard work, we finally have a theory that describes the world around us - the Standard Model - and we understand how to use the fundamental
parameters of that theory to describe the structure of that world. The Standard Model itself is a beautiful structure based on gauge symmetries. However, there is a not-so-beautiful aspect also: to fully specify the theory one need to give the values of 26 parameters - masses, mixing angles, coupling constants. These have no known symmetry and don't display any apparent logic. We expect that a more fundamental theory will give us insight into these parameters.

Given the success of the theory, we should be able to also understand what the world would look like if these parameters were modestly different.  In the neighborhood of their physical values, we should be able to understand what happens under variation in the parameters.

Performing this exercise leads us to the remarkable conclusion that the modest variation of a few parameters would drastically change the world as we know it. In some cases, atoms and nuclei no longer exist, and the universe would be sterile. Various aspects of this conclusion are discussed in \cite{agrawal, agrawal2, Damour, Jaffe:2008gd, Golowich, Barr:2007rd, Jeltema:1999na, Donoghue:2000ix, Meissner:2014pma, Epelbaum:2012iu, Hall:2014dfa, Yoo:2002vw}. Cosmology provides another constraint, with the well-known argument that if the cosmological constant were much different, matter would not clump into stars and planets \cite{Weinberg1, Weinberg2, Martel:1997vi, Tegmark:1997in, LindeMoriond, GarrigaV, Garriga2, DonoghueRandom, Banksanthropic, Bjorken}. We can refer to these as anthropic constraints \cite{Barrow:1988yia}. Although the word anthropic has its roots in the Greek word for human, these do not refer to us specifically but are really constraints on rather general physical properties of the world that we see. It takes careful science to delineate these physical constraints.

The anthropically allowed portion of parameter space appears to be a very small portion of what we would expect for the overall size of the parameter space. The world appears to be fine-tuned for atoms, nuclei, stars, complexity, etc. What do we make of this fact? It could be that the ultimate fundamental theory has a unique ground state that just happens to have all the right values for these parameters (i.e. good luck indeed!). Or perhaps there are many ground states with different parameters and the universe realizes domains with these different parameters (i.e. the multiverse). In the latter case, we should not be surprised that we find ourselves in a domain with the right properties. This reasoning provides a bottom-up motivation for multiverse theories.

Here come the caveats: Perhaps there are other islands in parameter space where enough complexity is found to lead to an interesting world, perhaps even with a form of life\footnote{An example of this phenomenon, with nearly degenerate $\Delta$ baryons and the proton, was identified in Ref. \cite{agrawal2}.}. Certainly, if the parameters were wildly different from ours, an exploration of such a situation would be much more difficult and much less reliable, and we cannot rule out the possibility of other islands of viable parameters. However, such situations do not appear to dominate the overall parameter space. This leaves the fundamental motivation unchanged. Even if some other possibilities exist, the viable neighborhood of our parameters appears to be puzzlingly small.

In addition, the structure of the world is not sensitive to all of the parameters of the Standard Model. If the mass of the top quark were modestly different, the structure of atoms and nuclei would be essentially unchanged. But for a handful of parameters (or more realistically combinations of parameters), it appears that the anthropic constraints apply.

Finally, it must be acknowledged that it is possible to be too extreme and attempt to claim that anthropic selection is the determining feature for most of the features of our world. Given how little we understand about potential theories with many ground states, the utility of this is not clear at present. There can also be a debate about exactly where the boundaries of the viable parameters are. This is a valid scientific discussion about the reliability of our present techniques, but in the big picture it is beside the point. In this review, I will allow the estimates of the precise constraints to be somewhat rough. The important point is that such boundaries in parameter space exist and that the viable region constitutes a very small part of the overall parameter space.

\subsection{Atoms and nuclei}

The complexity that we see in the world arises because we have many elements, which can be arranged in various ways. If we dig deeper into the fundamental theory, we see that having several elements relies on the feature that the up quark, the down quark and the electron have masses which are small compared to the QCD scale, and which have a particular ordering. That the fermion masses satisfy these properties is a bit strange, because the quark masses actually arise from the weak interaction, which is completely independent of QCD. They are the product of a Yukawa coupling and the Higgs vacuum expectation value (vev), and the vev is much larger ($v=246$~GeV) than the QCD scale. Roughly stated, the weak interactions must overlap with the strong interactions in order to have atomic structure.

In addition to the strong and weak interactions, electromagnetic effects play a role in determining the spectrum. We will see that a subtle dance of many features is needed in order to have the elements. In what follows, I will discuss individual masses or couplings, initially treating them as independent variables from each other and later recognizing that this independence may not be correct. However, this separate treatment allows at least a preliminary understanding of the constraints on the parameter space.

\subsubsection{The up quark mass}

The up quark and down quark masses are some of the more obscure parameters of the Standard Model. They are small on the scale of QCD, with the Particle Data Book \cite{RPP} quoting values of 2.1 MeV and 4.7 MeV respectively when defined at the running scale of 2 GeV. Yet they play an oversized role in the structure of our world. The near equality of the neutron and proton masses is not a fundamental symmetry of the Standard Model. Rather it appears as an accidental near-symmetry which occurs because both of these quark masses are so small that they are a small correction to the overall nucleon mass\footnote{The overall nucleon mass arises from the scale of QCD and is mostly independent of the quark masses.}. Moreover, the fact that the proton is stable is due to fact that the up quark is slightly less massive than the down quark.

If the up and down quark masses were equal the proton would be heavier than the neutron and hydrogen would not be stable. The down quark is required to be heavier in order to stabilize the proton. Recent lattice calculations \cite{Borsanyi:2014jba} put the electromagnetic effect at 1.0 MeV in favor of the proton and the the effect of $m_d-m_u$ at 2.4 MeV in favor of the neutron. Accepting these values, this implies that if the up quark mass were increased by just 1 MeV, the proton would be unstable.

If the up quark mass were yet slightly larger, other changes in the elements would appear. Eventually, even bound protons would be unstable to transitioning to a free neutron through electron capture. It would be a neutron world, and no atoms would exist. Using the average binding energy per nucleon of 10 MeV, and the effect of $m_d-m_u$ mentioned in the previous paragraph, one can estimate that this would occur if the up quark were 12 MeV heavier, with all other parameters held fixed. Since quark masses runs up to $m_t=1.7 \times 10^{5}$~MeV, it is clear that the allowed up-quark window is a tiny proportion of this range.

\subsubsection{The electron mass}

The atomic physics of the previous section can be adapted without much change to also constrain the electron mass. The neutron-proton mass difference is 1.29 MeV, and the electron mass is 0.511 MeV. If the electron were heavier than this 1.29 MeV, the hydrogen atom would not be stable but would decay into a neutron and a neutrino. Likewise when the mass increases much beyond 10 MeV, all the bound protons would combine with electrons to decay to neutral matter. Leptons have masses up to 3,500 times the electron mass. Atoms would only exist if the electron mass falls into a small part of this range.

\subsubsection{The fine structure constant}

Of the parameters of the Standard Model, the one that most likely has a good dynamical explanation is the fine structure constant. Within Grand Unified Theories \cite{GUTS}, if one postulates that the three gauge couplings of the Standard Model are subsets of a unified gauge coupling at high energy, the fine structure constant naturally comes close to the observed value. Yet within the Standard Model itself, this idea does not work exactly in practice. To make it work, new physics of a particular form\footnote{Supersymmetry at the weak scale appears to work.} at relatively low energies is required in order to modify the running couplings. However, we have do not have evidence for Grand Unified theories, nor (yet) of the needed low energy physics. Given this, we should at least consider the possibility that, like Kepler's Platonic solid construction, this elegant construction is not what Nature has chosen.

Electromagnetic effects by themselves would tend to make the proton heavier than the neutron. The simplest estimate would involve the energy contained in the electromagnetic field external to a charge distribution of a size $R\sim 1$~fm, $U= \frac12 \int d^3x ~E^2 \sim 3\alpha/5R \sim 0.88$~MeV. The electromagnetic interactions of the quark interior to the neutron and proton is more complicated but again favors the proton\footnote{The electromagnetic self-energies of the quarks themselves is somewhat ill-defined but also favors the up quark - and hence the proton - and is considered small.}. If the fundamental electric charge $e$ was twice as large, with other parameters held fixed, the proton would be about 1.6 MeV heavier than the neutron. This estimate again comes from the lattice calculation referred to above \cite{Borsanyi:2014jba}. This sets a rough bound on the viable range that says that the electric charge should not be more than 33\% larger than its physical value if hydrogen is to be stable.

\subsubsection{The down quark mass}

The considerations of the previous sections can also apply to the down quark mass. If it becomes slightly larger, by a few MeV, deuterium is no longer bound as the element has only a binding energy of 2.2 MeV. This becomes a problem for nucleosynthesis (see below) as the deuteron is a key step in the synthesis of the light elements. If it became yet larger, bound neutrons decay into free protons plus an electron and antineutrino. The world would be one of hydrogen, with very little chance for complexity. This happens when increasing the down quark by perhaps 10 MeV.

This same set of constraints occurs if we try to rescale all quark masses by a common factor, for example by increasing the value of the Higgs vev, keeping all their ratios fixed. Since $m_d/m_u >1$ eventually the neutron-proton mass difference becomes great enough. This was estimated in Ref. \cite{agrawal, agrawal2, Damour} to occur if all the masses are increased uniformly by about 65\%.

\subsubsection{The average of the light quark masses}

The above discussion of the neutron-proton mass difference yields constraints which apply mainly to $m_d-m_u$, so one might expect that if one held $m_d-m_u$ fixed there would be a large range for their sum. However, different physics enters to constrain the sum or average of these masses. The average $\hat{m} = (m_u+m_d)/2$ enters physics most importantly in the pion mass. Chiral physics predicts that the pion mass-squared is proportional to this quantity $m_\pi^2 \sim \hat{m}$. The vanishing of the pion mass at $\hat{m}=0 $ is a consequence of Goldstone's theorem for broken chiral symmetry in QCD. The dependence of the mass and couplings of other hadrons on $\hat{m}$ is quite small, but for the pion mass the dependence is dramatic.

The pion is the lightest hadron, and its mass most directly influences everyday physics through nuclear binding. The long range forces in the nuclei most responsible for binding depend on pion exchange. It is interesting to note that nuclear binding itself is a delicate balance. The average binding energy of 10 MeV per nucleon is very small on the typical scales of QCD, which typically are hundreds of MeV. The short range component appears to be largely repulsive. In the central nuclear potential, the long range attractive portion occurs in the two pion channel. It is often modeled by the exchange of a sigma meson, and we now know that the sigma is a strong coupled resonance of two pions. When modeled by meson exchange, the attractive and repulsive components would each be a large value closer to the typical QCD scale, but the sum produces a shallow potential with a much reduced binding energy. Single pion exchange is relevant for the light elements, but much less so for heavy elements because it is proportional to spin and isospin quantum numbers which average closer to zero in heavy nuclei.

Increasing the pion mass reduces the long range attractive component of the central nuclear potential. Because of the delicate balance described above, this can destroy nuclear binding of most elements relatively easily for even a modest mass increase. Estimates of Ref. \cite{Damour} using modern understanding of $\pi \pi$ scattering, following on earlier estimates of \cite{agrawal, agrawal2}, indicate that the pion mass increases even by 30\% the binding of heavy nuclei disappears. Estimates made using fundamental sigma fields are somewhat weaker, but this may due to the lack of modeling of the quark mass dependence of the fundamental sigma properties. It will be interesting in comparing the parameter dependence of the analytic potential of Ref. \cite{Damour} with the emerging numerical calculations using lattice methods \cite{Meissner:2014pma}. Preliminary comparisons are encouraging. Overall we conclude that the average mass of the light quarks also is tightly constrained.

\subsubsection{Nucleosynthesis}

The above constraints have focused on the existence of atoms and nuclei. Another set of considerations could be whether the elements get synthesized in the Universe. This is potentially quite a difficult study because, although we have mapped out the standard path of nucleosynthesis quite well, there could be alternate paths which also lead to a state with a sufficiently complex set of elements. However, there is no point in going overboard in this exercise. For our motivational purposes, it is again sufficient to note that fairly simple considerations of the neighborhood of our parameters indicate that our standard mechanisms for nucleosynthesis fall apart under quite modest variation in the parameters. This reinforces the sense of fine-tuning for our point in parameter space.

The most sensitive measure is the stability of the deuteron. Both the pathways of primordial nucleosynthesis and further synthesis in stars relies on deuterium as initial step. Removing it would require alternate pathways with quite different outcomes. This is interesting because the deuteron is just barely bound, by 2.2 MeV which is tiny on the QCD scale. Small changes in the quark masses readily unbind it. This is a variation of the nuclear/atomic considerations above, and simply provides a tighter constraint \cite{agrawal, agrawal2, Damour, Golowich}. Even yet tighter constraints can be obtained if one considers the triple alpha process, used in the generation of the heavier elements, to be essential \cite{Jeltema:1999na, Meissner:2014pma}.

The relative amounts of neutrons and protons produced primordially provides a constraint that is different in character from those described above. These relative amounts, which lead to relative amounts of hydrogen and helium, are determined by the strength of the weak interactions. The strength is determined mainly by the mass of the W boson, or equivalently by the Higgs vacuum expectation value. If the W mass is too light, then neutron decay happens rapidly and all the neutrons decay as they are falling out of thermal equilibrium, leading to a world of hydrogen only. At the other, more interesting, extreme if the W mass is much heavier, then the neutron and proton amounts get locked in earlier at a high temperature where their mass difference is irrelevant. This leads to an almost equal ratio of neutrons and proton and these get processed into dominantly Helium. In Ref \cite{Hall:2014dfa}, the need for surviving hydrogen has been converted into a constraint that the W mass (or Higgs vev) is not more than 5 times the observed value.

\subsubsection{Correlated constraints}

Considering the viable range of a single parameter clearly understates the available parameter space. For example, if we raise the up, down and electron masses keeping the mass differences $m_n-m_p$ and $m_n-m_p-m_e$ unchanged, then some of the constraints listed above no longer apply and so the available parameter space is enlarged. However, eventually one runs into a boundary due to the average of the up and down quark masses. In addition, we do not even know which parameter to vary. Is it the Higgs vev, which changes the quark and lepton masses and the W mass in parallel, the most important variable? Or should all the
fundamental parameters of the Standard Model vary independently? One can try looking for extreme situations that leave the essential physics unchanged \cite{Harnik:2006vj, Clavelli}. These are some of the caveats about the exercise described above.

However, the generic conclusions remain unchanged. There are relatively tight constraints on about 5 combinations of parameters and small changes in these combinations lead to major changes in the structure of the world. Briefly stated, it is that the weak interactions must overlap with the strong interactions. The light quarks and the electron masses are the product of the Higgs sector of the Standard Model, and these are constrained to be in a small window far below the QCD scale. On the other hand the W mass is larger than the QCD scale, and this is relevant for nucleosynthesis. The allowed parameter space seems to very small and the world as we know it seems highly fine-tuned for the existence of atoms and nuclei.

I have spent this much space describing the atomic constraints because they provide a unique form of motivation for multiverse theories. One has to see the many constraints in order appreciate just how very small a portion of parameter space is available which leads to atomic structure. Moreover, there seems to be no possibility of a dynamical mechanism to generate parameters in this small window. Unlike the cosmological constant or the Higgs vacuum expectation value, for which dynamical mechanism are at least sought to explain their small values, even theories attempting to describe the Yukawa couplings would have to be just plain lucky to have the outcome fall in just the right window to explain atoms and nuclei.

\subsection{The universe and the cosmological constant}

The most well known anthropic constraint is that of the cosmological constant \cite{Weinberg1, Weinberg2, Martel:1997vi}. Here the physical constraint is on the gravitational clumping of matter into stars and planets. If the cosmological constant is positive and too large, the universe expands so rapidly that clumping does not occur. If it is negative and too large, the universe will have collapsed before clumping occurs. While again the boundaries of the allowed region are not exceptionally precise, it appears that the cosmological constant should be within two orders of magnitude of its observed value\footnote{The constraints on the cosmological constant are also correlated with amplitude of cosmological density fluctuations\cite{Tegmark:1997in}} - and perhaps even more tightly constrained.

This constraint is exceptionally tight. The cosmological constant is the energy density of the ground state of the theory. The observed value is $\Lambda_0= 2.4 \times 10^{-47} ~ {\rm GeV}^4 = (2.2 \times 10^{-3} {\rm eV})^4$. There are very many contributions to the vacuum energy. However, these are all expected to carry energy scales which are much larger than $(10^{-3} {\rm eV})^4$. Zero-point energies of quantized fields are formally divergent but hopefully either cancel or are made finite at some high energy. The Higgs potential carries energy densities of order $10^{51}\Lambda_0$. The strong interactions bring in contributions of order $10^{47}\Lambda_0$. Many of these contributions are difficult to calculate precisely, but their order of magnitude should be correct. One needs to have all such contributions cancel to at least 50 or so orders of magnitude.

I would like to explain in some detail one contribution to the cosmological constant that can be calculated with great reliability. It has not been described in the literature before to the best of my knowledge\footnote{However, it does appear as an exercise in \cite{DSM}}. It also is a good illustration of how difficult it would be to adjust the parameters of the Standard Model to bring the cosmological constant in line. We will see that we need to specify the 41st digit of the up quark mass if we are trying to adjust the parameters to give the correct cosmological constant.

The contribution comes from the shift in the vacuum energy caused by the explicit breaking of chiral symmetry, and the reliability of the calculation comes from the use of symmetry in chiral perturbation theory\cite{DSM, Gasser}. First consider two-flavor QCD without the up and down quark masses, but with external scalar and pseudoscalar currents $s(x),~p(x)$,
\begin{equation}
{\cal L} = -\frac14 F^a_{\mu\nu}F^{a\mu\nu} + \bar{\psi} i \slashed{D}\psi - \psi_L (s+ip)\psi_R - {\bar{\psi}}_R (s-ip)\psi_L
\end{equation}
Here $\psi$ is an $SU(2)$ doublet field of the up and down quarks. This system has an exact $SU(2)_L\times SU(2)_R$ chiral symmetry
\begin{eqnarray}
\psi_L \to L \psi_L~~~~~~~~~\psi_R \to R \psi_R~~~~~~~~~(s+ip) \to L(s+ip)R^\dagger
\end{eqnarray}
where $L$ is an element of $SU(2)_L$ and $R$ is an element of $SU(2)_R$. This version of massless QCD undergoes dynamical symmetry breaking, with pions being the Goldstone bosons. The resulting low energy effective Lagrangian for the pion manifests the $SU(2)_L\times SU(2)_R$ symmetry and can be expanded in an energy expansion. At lowest order, this results in an effective Lagrangian (in conventional notation \cite{Gasser})
\begin{equation}
{\cal L}_2 = \frac{F_\pi^2}{4}\left[ Tr\left(\partial_\mu U\partial^\mu U^\dagger\right)+ Tr \left(\chi U^\dagger + U\chi^\dagger\right)\right]
\end{equation}
where the $2\times 2$ matrix $U$ contains the pion fields $\pi^i,~~{i=1,2,3}$
\begin{equation}
U = \exp {i\frac{\tau \cdot \pi(x)}{F_\pi}} \ \  ,
\end{equation}
where $\tau^i$ are Pauli matrices and the the external sources are contained in $\chi = 2 B_0 (s+ip)$ with $B_0$ being a constant of dimension (mass)$^1$ and $F_\pi$ is the pion decay constant $F_\pi=92$~MeV. This Lagrangian displays the exact chiral symmetry
\begin{equation}
U \to L U R^\dagger ~~~~~~~~~~(s+ip) \to L(s+ip)R^\dagger  \ \ .
\end{equation}
For more on the construction of effective chiral Lagrangians, see Ref. \cite{DSM}.

Real QCD including quark masses is obtained by replacing the external fields by the quark mass matrix,

\[
s
=
\begin{bmatrix}
    m_u & 0  \\
    0 & m_d
\end{bmatrix}
~~~~,~~~~~p=0
\]
In this case, the pions pick up a small mass
\begin{equation}
m_\pi^2 = B_0 (m_u+m_d) \ \  .
\end{equation}
which is found by expanding the Lagrangian to second order in the pion field. However, for our purposes the effective Lagrangian also yields
a contribution to the vacuum energy of the form
\begin{equation}
\Lambda_m = -\langle 0| {\cal L}_2 |0\rangle = - F_\pi^2 B_0 (m_u+m_d) = - F_\pi^2 m^2_\pi   \ \ .
\end{equation}
This relation can also be gotten without the construction of the effective Lagrangian by using first order perturbation theory and the soft pion theorem \cite{DSM}
\begin{equation}
\Lambda_m = \langle 0| m_u \bar{\psi}_u\psi_u + m_d\bar{\psi}_d\psi_d |0\rangle = - F_\pi^2 \langle \pi| m_u \bar{\psi}_u\psi_u + m_d\bar{\psi}_d\psi_d |\pi\rangle  = - F_\pi^2 m^2_\pi   \ \  .
\end{equation}
However, I have used the effective Lagrangian approach because it makes clear that there is no possible compensating term linear in the the quark masses\footnote{The first unknown contribution comes at second order with a Lagrangian of the form $Tr (\chi\chi^\dagger)$ which contributes to the vacuum energy but not to the phenomenology of pions \cite{Gasser}. However, second order contributions are much smaller and have a different functional dependence on the masses.}.

This contribution to the vacuum energy is precisely known,
\begin{equation}
\Lambda_m = 1.5 \times 10^{8}~{\rm MeV}^4 = 0.63 \times  10^{43} \Lambda_0  ~~~~~~.
\end{equation}
If we consider a situation where the quark mass parameters were slightly different, this would be expressed as
\begin{equation}
\Lambda_m  =  0.63 \times 10^{43} \Lambda_0 \frac{(m_u+m_d)}{(m_u+m_d)_{phys}}  ~~~~~~.
\end{equation}
Because of the large multiplier, if one holds all the other parameters of the Standard Model fixed, a change of the up quark mass in its 41st digit would produce a change in the cosmological constant outside of the anthropically allowed range. Of course, this should not be treated as a real bound on the up quark mass variation, because small changes in other parameters could compensate for this shift in $\Lambda$. There are too many parameters that also contribute enormously to $\Lambda$ which could have a potentially correlated variations keeping $\Lambda$ fixed. However because the calculation is so well controlled it does illustrate the degree of fine-tuning required and illustrates the futility of thinking that some feature of the Standard Model could lead a vanishing contribution to $\Lambda$.

Space considerations of this review do not allow a full discussion of other ties between physical properties and anthropic constraints. Some others include neutrinos \cite{Tegmark:2003ug}, dark matter \cite{Freivogel:2008qc, Hellerman:2005yi, Bousso:2009ks, Bousso:2013rda} and even the dimension of spacetime \cite{Tegmark:1997jg}. Unlike most of the discussion above where a well defined theory - the Standard Model - underlies the discussion of the variation of parameters, the fundamental theory for these latter considerations is not known so that usefulness of these constraints are less clear.

\section{NATURALNESS VS MULTIVERSE AND THE FINE-TUNING PROBLEMS OF PARTICLE PHYSICS}

In particle physics, the phrase ``fine-tuning'' generally has a different meaning from that used in the previous section. In that section, it was noted that only a small range of parameter space is compatible with atoms, nuclei, stars, etc.. This is fine-tuning of parameters in order to allow a set of physical properties. However, more commonly in particle physics fine-tuning is used to describe the situation where a parameter is observed to be much smaller than its expected ``natural'' size \cite{'tHooft:1979bh, Dine:2015xga}. The three fine-tuning problems in this sense are the cosmological constant problem, the Higgs vev problem and the strong CP problem. Of course, the two meanings of fine-tuning can overlap, as they do in the first two of these problems. That overlap is the source of much of the interest in the multiverse.

The conventional response to perceived fine-tuning is to look for a physical mechanism to make this occurrence technically natural. Naturalness implies that the various contributions to a parameter, both classical and through quantum radiative corrections, are of the same order of magnitude so that no delicate cancelations are present. For example, a classical contribution to the cosmological constant comes from the minimum energy of the Higgs potential treated as bare parameters, and quantum effects could be zero-point energies or radiative corrections, among other effects. While logarithmic divergences in radiative corrections are technically infinite, these are not viewed as barriers to naturalness because, with any reasonable cutoff, the logarithm is not very large.

Both the ideas of naturalness and of the multiverse do not constitute theories in themselves but serve as motivations for new theories. Using naturalness as a motivation, one requires new particles and interactions beyond the Standard Model which make the full theory technically natural. Although the cosmological constant is the greatest problem for naturalness, the focus at the moment is on the Higgs vev problem because naturalness predicts that new physics will be discovered at the LHC. In contrast, using the multiverse as a motivation, one requires the development of models with a great number of ground states in order to allow the fine-tuning to be understood as a selection effect. If you have enough ground states, neither the cosmological constant nor the Higgs vev are considered as problems, although the strong CP problem remains.

\subsection{The cosmological constant}

The idea of naturalness appears to fail for the cosmological constant. Applied to the cosmological constant, naturalness would imply new particles and interactions at the scale $10^{-3}$ eV. Although such a model has been proposed \cite{Sundrum:2003jq}, to the best of our knowledge Nature does not take this path and technical naturalness fails. In the opinion of many, the multiverse remains the best available explanation for the cosmological constant problem.

\subsection{The Higgs vacuum expectation value}
The jury is still out for natural theories for the Higgs vacuum expectation value (vev). There are potentially two naturalness problems for the weak scale. One focusses on the quadratic divergences in the Higgs vev that occur when the Standard Model is treated in isolation.  When quadratic divergences are treated with a cutoff, these quickly become very large as the cutoff is raised and one requires a large fine-tuning to compensate. This is used to argue against theories with quadratic divergences, such as the Standard Model, as fundamental theories in isolation\footnote{This version of fine-tuning has been challenged \cite{Shaposhnikov:2008xi, Lynn:2011aa, Lynn:2015ala, Aoki:2012xs} as being incorrect, and there may be some merit to these arguments.}. This naturalness problem requires new physics at a TeV so that the known cutoff dependence does not get too large.

The second version of the naturalness problem (or the fine-tuning problem) occurs when the Standard Model particles also participate in other interactions with a larger fundamental scale, for example Grand Unified Theories. In this case, radiative corrections with the new interactions would tend to bring that larger scale also into the Higgs potential, raising the Higgs vev to a larger scale. It is not just divergent effects that are at issue here, even finite quantum corrections could require large fine-tuning. Again, some forms of new physics at the TeV scale could solve this. At the time of this writing, the LHC has not found evidence of such new physics even though it has pushed significantly into the energy range where it should occur. However, further experiments continue pushing deeper into the realm of possible new physics and we all eagerly await the results.

The work of Refs. \cite{agrawal,agrawal2} were the first to point out that the multiverse and anthropic selection of atoms could potentially account for the Higgs vacuum expectation value. In this reference, for simplicity all parameters except the vev were held fixed, so that quark mass ratios were treated as constants. This yields a particular slice through the atomic constraints. It is clear that this assumption can be relaxed and subsequent work have done so \cite{Damour,  Donoghue:2009me, Hall:2007ja}. Anthropic selection of the overall weak scale remains a possibility. As with the cosmological constant problem, this option will gain in interest if more conventionally natural solutions are not uncovered.

\subsection{The strong CP problem}

The multiverse idea fails to resolve the strong CP problem \cite{Donoghue:2003vs, Dine1, Ubaldi:2008nf}. Stated more positively, it implies that this is the only one of the three big naturalness problems for which a dynamical explanation is required. The strong CP problem refers to the CP and T violating term proportional to $\theta \epsilon^{\mu\nu\alpha\beta}F_{\mu\nu}F_{\alpha\beta}$ that must be included in the QCD action. Because CP is violated in the Standard Model, there is no good reason to set $\theta$ to zero, and the effect of $\theta$ has an additive contribution, expected to be of order unity, from the phases in the Yukawa couplings that generate other forms of CP violation. Yet experiments on the neutron's electric dipole moment require the coefficient of this effect to be smaller than $10^{-10}$. If $\theta$ were many orders of magnitude larger, there would not be any significant change in the structure of the world. So the multiverse cannot be used to invoke anthropic selection, and leaves this problem for a dynamical explanation. There is in fact a good dynamical explantation involving the Peccei-Quinn symmetry \cite{Peccei:1977hh} and the axion \cite{Wilczek:1977pj, Weinberg:1977ma}. Both naturalness and the multiverse serve as motivations for searching for the axion.

\subsection{An axion multiverse}

It is worth noting that if the axion is indeed the solution to the strong CP problem, the multiverse idea may enter quite naturally in a slightly different way \cite{Hertzberg:2008wr, Wilczek:2013lra, D'Eramo:2014rna}. In the standard picture of the early universe, the initial value of the axion field is not fixed but would be randomized by quantum and/or thermal fluctuations. In the presence of inflation, the regions of different initial values would be inflated such that different initial patches would become causally disconnected domains at late times. We would live in one of these, but other places in the greater multiverse would have different initial axion values. Because the initial value of the axion determines the extent to which the axion contributes to dark matter, the different domains would also have different amounts of dark matter. This realization also plays a role in axion phenomenology. Naturalness has been used to rule out axion theories with a large value of the axion decay constant, because in this case a random initial value of the axion would lead to too much dark matter. However, since too much dark matter can also have a negative consequence on the evolution of the Universe, there is in fact an anthropic selection of the possible domains which allows an allowed anthropic window with large decay constant \cite{Hertzberg:2008wr, Wilczek:2013lra, D'Eramo:2014rna}.

The axion multiverse shows that multiverse behavior can show up even with completely conventional physics. Here we are not talking about hypothetical string vacua, but rather a very standard new particle. If axions with the right properties are the source of dark matter, then the amount of dark matter is an accident of the history of our particular patch of the universe and attempts to predict this quantity is no more useful than attempts to predict the radius of the Earth's orbit. The axion also is an interesting example of how the multiverse could arise from a continuous variation in the properties if the universe.

\section{PHYSICAL MECHANISMS - MODELS FROM THE TOP DOWN}

The multiverse idea would have little relevance unless there were physical theories that potentially lead to this feature. As mentioned in the introduction, the most challenging ingredient is the requirement of multiple ground states. Sociologically, the multiverse idea got a major boost when it was argued that
string theory had this property. However, even string theory cannot yet be counted as a complete model because we do not have enough control over the theory to make even statistical predictions.

\subsection{Many states}

It was originally argued that string theory would have a unique ground state and that this would be the signal that it was the correct theory. It would indeed be impressive if such a unique state were identified and all of the masses of the up quark, the down quark etc agreed with experiment. However, it is looking less likely that this will occur. The number of string ground states seems enormous \cite{douglas, Ashok, Denef:2004ze, Bousso, Susskind} - some counts estimate $10^{500}$ - and we do not have a principle to select a unique one out of the multitude. The string ground states are described not only by the way that the extra dimensions are compactified in order to leave our four dimensions, but also by the way that the field components are arranged within these internal spaces. Spaces with different amounts of flux wrapped around the internal dimensions have different low energy properties.

An interesting feature to keep in mind is that changing one of the fluxes by one unit can be a large change in the properties. The fact that there can be a near-continuum of values of the parameters is not because the flux quanta are each individually small, but rather that there are so many ways to combine up the fluxes that one saturates almost every possible result \cite{Bousso}.

It is not only in string theory that one can obtain multiple ground states. For example, an old suggestion related to the cosmological constant \cite{Brown1, Brown2} invokes the possibility of three-form potentials $A_{\alpha\beta\gamma}$ with four-form field strengths $F_{\alpha\beta\gamma\delta}$. In four dimensions, the four-form Lagrangian $F_{\alpha\beta\gamma\delta}F^{\alpha\beta\gamma\delta}$ leads to equations of motion which fix the field strength to be a constant, but which can take on different quantized values (and tunnel between such values) if coupled up to a charge. The four-form action then is an extra positive contribution to the cosmological constant which in principle can take on arbitrary values. Using higher dimension operators with the four-form coupled to other fields can then also change other particle properties and couplings.

In addition, if there are large numbers of fields, Dvali and Vilenkin \cite{DV1, DV2} have identified field theoretic ways to have large number of vacua. Both the mechanisms of \cite{Brown1, Brown2} and \cite{DV1, DV2} also illustrate a caveat to theories with many vacua. Both also have dynamical mechanisms at work that populate preferentially some states over others, depending either on the past history of our domain or on the density of states of the vacua.

The axion example described above also illustrates the possibility of a continuous variation of the parameters across the universe. In the axion case, the initial value of the axion is really a continuous field variation in the early universe. Our patch of the universe is then taken from a very small segment of this field and inflated so much that it looks uniform across the sky today. Because the initial axion field value determines the amount of dark matter that we see, other elements of the greater universe outside of our local patch would see a continuous variation of the abundance of dark matter. If desired this mechanism could be applied to other properties also. However, the axion example also illustrates an extra feature of models with a continuous variation of the parameters - there will be a light field associated with such variation. If a parameter has spatial or temporal variation, it is a field. For the spatial variation to span a large region of space, that field must be light. However such light fields can potentially evade present attempts at detection, as indeed the axion has so far.

We see that there is considerable room for model building of theories with multiple ground states. This has not been a priority for the community, which has probably been a wise choice at the moment. However, it is useful to recognize that multiple ground states is a physical possibility and landscapes can exist outside of string theory.

\subsection{Populating the multiverse}

We also need to know that there is a mechanism for populating the different vacua in different regions of the universe. Since this process would occur in the very early universe - before or during inflation - we can let our imaginations run wild and be confident that some such mechanism can always be found. Nevertheless, some mechanisms are already known\cite{Bousso, Donoghue:2003vs, Brown1, Brown2, wilczek, Escoda:2003fa, Linde1, Linde2, Garriga}. At high temperatures all states get populated, and causally disconnected regions would settle in different ground states. Tunnelling between different discrete ground states has been explored using the four-form field strengths \cite{Brown1, Brown2} and this can be adapted to string vacua \cite{Bousso, Donoghue:2003vs}. In the initial four-form problem, the step size for the changes in the cosmological constant are treated as very small, by taking the associated charge to be very tiny. However, in the string vacua case, the step sizes for any changes would be expected to be large, as one is breaking a flux factor in a highly compact internal dimension. But with enough inflation (generically eternal) the different ground states will all be sampled.

\section{TESTING THE MULTIVERSE}

Now comes the weakest part of this review. There are several problems.  We do not have a concrete multiverse theory with which we can make predictions. The existence of distant spatially separated domains is likely not directly testable because they are likely causally disconnected from us. The general nature of a multiverse theory would even make it hard to make specific predictions because the parameters of the theory are by definition not uniquely fixed.

We don't generally have to test every prediction of a given theory. For example, in the Standard Model, it is unlikely that we will ever be able to test the dramatic prediction that the baryon anomaly predicts that the proton is not stable, because the predicted lifetime is about $10^{130}$ years. Nevertheless, there must be enough predictions of a theory that are tested that we can trust it in its expected range of validity. The Standard Model has passed such tests. The multiverse will have difficulty here.

One possibility is that the parameters of the Standard Model could still provide a test of an underlying theory, although that test would likely be statistical in nature. For example, we have 6 quark masses, 3 charged lepton masses, two neutrino mass differences, and the weak mixing angles of the quarks and leptons. All of these come from the Yukawa couplings of the theory. In a multiverse theory, we should not expect that these couplings are uniquely predicted. However, the underlying theory would presumable tell us how they are distributed, i.e. their measure. Are the masses in the theory uniformly distributed on a linear scale or perhaps on a logarithmic scale? The distribution of Yukawa couplings would also influence the size of the weak mixing angles, which arise from diagonalizing the original Yukawa matrices.

The experimental measure of the quark and charged lepton masses turns out to be rather striking \cite{weight, Donoghue:2005cf, Donoghue:2009me}. If we treat all masses as independent, they appear to have a {\em scale invariant distribution}. That is, they are randomly distributed on a logarithmic scale. Quantitatively, if we propose that the weighting of the distribution is $dm/m^\delta$, the fit value of the exponent is $\delta =1.02 \pm 0.08$, with $\delta =1 $ corresponding to a scale invariant measure. The exponent is relatively well determined despite the small number of masses, and the quality of the fit is excellent. The scale invariant distribution also naturally leads to a hierarchy of the quark mixing angles. The neutrino masses are considerably lighter and would not fit this pattern well if they were generated in an identical fashion. However, even in more standard settings the light masses are generally treated as evidence of other mechanisms of mass generation at work, such as the seesaw mechanism \cite{seesaw}. However, explorations of this possibility indicate that even for neutrinos it is plausible that randomness generates the masses and mixing angles \cite{Donoghue:2005cf, Hall}.

Of course there are caveats here. The complete independence of the Yukawa couplings would not naturally account for the generation structure of the weak doublets. In addition, the anthropic constraints on the up, down and electron were not accounted for\footnote{However, simply removing the $u, ~d,~e$ from the fit still leaves the scale invariant form preferred although with a larger uncertainty.}. To fully address these and other caveats, one needs a complete controllable multiverse underlying theory with multiple ground states. This would allow one to address, from the top down, the correlations between parameters which appears in the solutions.

Nevertheless, it is clear from the preliminary investigation of the measure for masses that there will be a statistical test of an underlying theory. It may have a somewhat different form than that of the independent mass hypothesis above, but the statistical power is present to allow a discrimination of theories.  If that theory predicted a flat distribution of masses on a linear scale it would be clearly incorrect. So at least in this statistical sense, it would be possible to falsify a given multiverse theory.

In a particular variant of string phenomenology it is possible to connect the observed Yukawa couplings back to aspects of string theory. In brane-world realizations \cite{ibw, Ibanez1, Ibanez4}, the Yukawa couplings ($y$) arise from non-perturbative effects which contain an exponential dependence on the area ($A$) of overlap of different branes,
\begin{equation}
y \sim e^{-c A}
\end{equation}
With this relation, the scale invariant distribution of Yukawa couplings corresponds to a flat distribution of areas (i.e. random on a linear scale). If this flat distribution could be derived in the brane-world scenario it would pass the Yukawa measure test.

Other suggestions for tests have also been put forth. For example, it has been suggested that perhaps one could find evidence of different domains by observing the CMB \cite{Kleban, Chang:2008gj} . If the amount of inflation is just small enough to barely separate the domains, the domain walls could be visible in the temperature fluctuations. If this occurred it would be a very dramatic and direct piece of evidence of other domains. However, it is not a necessary consequence of a multiverse. Most inflationary pictures produce more inflation that this, and the domain boundaries would be far removed from the CMB surface of last scatter. It would take luck to have just enough inflation to allow this effect to be visible.

If the variation of parameters is continuous rather than discrete, there could potentially be an observation of a spatial variation across the observed universe \cite{Webb:2010hc, Berengut:2010yu, Uzan:2010pm, Barrow:2002db, Damour:2011fa}. Such variations are being searched for independent of this motivation, but become even more interesting because of it.

It likely that the cosmological constant would be the most sensitive parameter \cite{Donoghue:2001cs}. This is due to the large cancellation that appears to be needed to get the small cosmological constant. As described above, a variation of the up quark mass of one part in $10^{43}$ would lead to an enormous variation of the cosmological constant across the sky. The most sensitive tests of this appear to be the parameters of the Cosmic Microwave Background (CMB) temperature fluctuations, which constitute the longest lever arm for experimental observations of spatial variations. Although an unexpected directional dependence in the power is seen in the data, it does not appear to be due to a dipole in the fit parameters describing the relative contributions of the fundamental parameters \cite{Evan}.

The multiverse also motivates new classes of theories that themselves could be tested more directly. An example is the idea of split supersymmetry \cite{ArkaniHamed:2004fb, Giudice:2004tc, ArkaniHamed:2005yv}. In a supersymmetric theory with the possibility of variable masses, if one wants to keep the Higgs vev light for atomic selection, it might be advantageous to have certain of the extra SUSY particles be light also. Other of the supersymmetric particles could then still be heavy. (An analogy is in the quark masses - there is an anthropic selection for the up and down quarks to be light, but the top and bottom can be heavier because there is no selection for them.) This idea is directly testable at the LHC, as it predicts that some supersymmetric particles would be found there but not others.

This is probably also the appropriate place to mention the contentious issue of developing a measure for the multiverse in the context of inflation \cite{mediocrity, GarrigaPrescription, Freivogel:2011eg, Bousso:2010id, Garriga:2012bc}. This is an attempt to assess how likely our patch of the universe could be in a universe that has other patches that continue to inflate. This is surprisingly subtle, as defining probabilities depends on whether one averages over volumes at one time slice or follows back worldlines to earlier times. From the view of particle physics a solution is not crucial for discussing what happens in our patch, as long a such patches exist at all. However, the inflationary measure problem is of conceptual interest for the development of a more complete multiverse theory.

\section{SUMMARY}

Given the physical possibility of theories with very many ground states, one of the great questions becomes ``Universe or Multiverse?''\footnote{Note also that a book by this name \cite{Carr:2007zzb} exists and provides the curious reader with much more to think about.}. Does the fundamental theory have a single ground state leading to our world or rather does it have many possible ground states of which we find ourselves in a domain with favorable parameters?

The existence of fine-tuning for stars, atoms and nuclei favors the multiverse option. We do not appear to live at a random point in parameter space, but at a special one allowing these features. One may judge for oneself how likely the good luck would have to be in order for a theory with a unique ground state to have parameters in the neighborhood of our special point\footnote{Browsing the internet on this topic will also reveal claims that fine-tuning reveals evidence of Deistic design, although this is not a scientific response. However, even this would not resolve the issue. Even the theologically inclined would need to question whether the Creator designed a rigid structure with only one option or a flexible one where the natural evolution would lead to life somewhere in the multiverse. There is a science question here. } . In particular the exponential fine-tuning needed for the cosmological constant seems to have no technically natural explanation, and is a strong motivation for a multiverse explanation.

It does seem that multiple ground states is a physical possibility. However, we have no complete theory that allows us to make even statistical predictions.

The paucity of experimental tests does not mean that multiverse theories are wrong. It would be unscientific to exclude theories with multiple ground states from consideration - they could be correct. However, it does mean that we need to be more modest in what we can expect from the study of the fundamental interactions. There may be questions that we cannot possibly answer about the origin of our theories. In the meantime, there are still great problems to address about the structure of Nature, such as the form of dark matter, the mechanism of baryogenesis and the combination of quantum mechanics with general relativity. Perhaps a satisfying unique theory with all the right properties will emerge, and we will be content to forget about the multiverse. But we have to recognize that we may not be able to construct such a completely satisfying theory.

Even more conventional theories may have have insurmountable obstacles to complete testing of the theory. Physics is an experimental science and there are sociological limitations to pushing exploration to ever higher energies. Full exploration of the Planck scale may never be possible and the best that we may hope for is an occasional and limited test sensitive to all the rich physics that we expect at that scale.  Multiverse theories may have different obstacles. The inherent limitations of testing multiverse theories will prove to be a barrier to full knowledge of the origin of the fundamental interactions if this is the solution that Nature has chosen.

However, as always, more work is needed. We are far from complete in our exploration of either conventional theories or multiverse theories. It is also possible that a very satisfying multiverse theory will be developed, and will explain presently connections between the different aspects of the world which we do not understand. In the meantime, both naturalness and the multiverse can play the role of motivations for new theories. In the case of naturalness, the cosmological constant and the Higgs vev are the greatest puzzles. We have developed a large range of theories to solve the naturalness puzzles of the Higgs vev, and are actively testing them at the LHC. Using the multiverse as motivation points not towards these, but rather to the strong CP problem as the greatest puzzle requiring a dynamical solution. Searches for the axion as the solution to this puzzle are also underway.

In Voltaire's philosophical fable {\em Candide ou l'Optimisme}  \cite{Voltaire} the title character and Pangloss, the professor of ``metaphysico-theologo-cosmolonigology'',  debate the question of whether this is the ``best of all possible worlds''. After examining the evil in the world, they give a negative assessment, yet retire to ``cultivate their garden''. Perhaps we have the physically equivalent question. We have seen that we are close to the best in terms of atoms, nuclei and stars, and have suggested that there could be other worlds. Our evidence is not clear yet. Yet the second part of the title - {\em l'Optimisme} - still should apply. We are in the early days of investigations of such theories and perhaps if we cultivate them some fruit will come. While we may run into barriers, it is still time to be optimistic and see where these theories lead.

\section*{ACKNOWLEDGMENTS}
Over the years, the author has gratefully received support for work related to this topic from the Foundational Questions Institute, the U.S. National Science Foundation (currently grant PHY-1520292) and the John Templeton Foundation. He also thanks very many colleagues for many lively discussions.

%

\end{document}